\pdfoutput=1
\documentclass[journal,twoside,web]{ieeecolor}
\usepackage{generic}
\usepackage{cite}
\usepackage{amsmath,amssymb,amsfonts}
\usepackage{algorithmic}
\usepackage{graphicx}
\usepackage{multirow}
\usepackage{subcaption}
\usepackage{algorithm,algorithmic}
\usepackage{hyperref}
\usepackage{booktabs}
\hypersetup{hidelinks}
\usepackage{textcomp}
\usepackage{caption}
\def\BibTeX{{\rm B\kern-.05em{\sc i\kern-.025em b}\kern-.08em
    T\kern-.1667em\lower.7ex\hbox{E}\kern-.125emX}}
\begin{document}
\title{A Dynamic Domain Adaptation Deep Learning Network for EEG-based Motor Imagery Classification}
\author{Jie Jiao, Meiyan Xu, Qingqing chen, Hefang Zhou, Wangliang Zhou,  Second B. Author, and Third C. Author Jr., \IEEEmembership{Member, IEEE}
\thanks{This paragraph of the first footnote will contain the date on 
which you submitted your paper for review. It will also contain support 
information, including sponsor and financial support acknowledgment. For 
example, ``This work was supported in part by the U.S. Department of 
Commerce under Grant 123456.'' }
\thanks{The next few paragraphs should contain 
the authors' current affiliations, including current address and e-mail. For 
example, First A. Author is with the National Institute of Standards and 
Technology, Boulder, CO 80305 USA (e-mail: author@boulder.nist.gov). }
\thanks{Second B. Author Jr. was with Rice University, Houston, TX 77005 USA. He is 
now with the Department of Physics, Colorado State University, Fort Collins, 
CO 80523 USA (e-mail: author@lamar.colostate.edu).}
\thanks{Third C. Author is with 
the Electrical Engineering Department, University of Colorado, Boulder, CO 
80309 USA, on leave from the National Research Institute for Metals, 
Tsukuba, Japan (e-mail: author@nrim.go.jp).}}

\maketitle

\begin{abstract}
There is a correlation between adjacent channels of electroencephalogram (EEG), and how to represent this correlation is a problem that is currently being explored. In addition, due to inter-individual differences in EEG signals, this discrepancy results in new subjects needing to spend a amount of calibration time for EEG-based motor imagery brain-computer interface. In order to solve the above problems, we propose a Dynamic Domain Adaptation Based Deep Learning Network. First, the EEG data is mapped to the three-dimensional geometric space, and its temporal-spatial features are learned through the 3D convolution module, and then the spatial-channel attention mechanism is used to strengthen the features, and the final convolution module can further learn the spatial-temporal information of the features. Finally, to account for inter-subject and cross-session differences, we employ a dynamic domain-adaptive strategy, the distance between features is reduced by introducing a Maximum Mean Discrepancy loss function, and the classification layer is fine-tuned by using part of the target domain data. We verify the performance of the proposed method on BCI competition IV 2a and OpenBMI datasets. Under the intra-subject experiment, the accuracy rates of 70.42$\pm$12.44 and 73.91$\pm$11.28 were achieved on the OpenBMI and BCIC IV 2a datasets.
\end{abstract}

\begin{IEEEkeywords}
 Attention, Brain Machine Interface, Deep Learning, Domain Adaptation Learning, Motor Imagery
\end{IEEEkeywords}

\section{Introduction}
\label{sec:introduction}
\IEEEPARstart{R}{ecently}, the brain-computer interface (BCI) has become one of the hottest research topics with wide-ranging applications in the field of medical health and rehabilitation training.
Motor Imagery (MI) is the cognitive process of imagining and is typically defined as imagining a part of the body moving without actually moving that part \cite{kwak2023subject}. Thereby this causes underlying changes in the cerebral cortex that can be observed in electroencephalogram (EEG) signals \cite{zhang2023motor}. Decoding the potential changes in the cerebral cortex caused by MI can transform biological signals into control commands for external devices, thus enabling BCI applications. Practical applications of the BCI technology include controlling device movement \cite{ai2023bci}, rehabilitation of stroke patients \cite{al2021eeg}, and entertainment for healthy individuals \cite{amini2023designing} and so on, so the correct classification of EEG signals has become a key issue in our research.

In an effort to use EEG signal-based MI classification, many machine learning methods based on EEG have been proposed. For example, using band-pass filter and common spatial pattern (CSP) algorithm to extract optimal spatial features, and utilized classification algorithm to classify CSP features \cite{fang2022feature}, applying the fast Fourier transform and CSP algorithm to extract MI features, and then employed three different classifiers to classify the transformed EEG data \cite{ko2019multimodal}. However, traditional frameworks require the process of feature extraction and feature classification to be separated, which not only requires a huge workload but also may lead to bias. With the rapid development of high-performance computing equipment, deep learning is widely used, which can learn features directly from data, also known as "end-to-end" \cite{zhu2022deep}. EEGNet \cite{lawhern2018eegnet} is a general deep learning framework that uses three convolutional layers to extract temporal and spatial patterns from EEG data and achieves good performance across multiple experimental paradigms. FBCNet \cite{mane2021fbcnet} constructs a multi-frequency band network structure to encode the spectral-spatial discriminant information related to MI. In addition, to fully utilize the features of various dimensions of EEG, people implemented a new 3D representation of EEG \cite{gao2023sft}, a multi-branch 3D convolutional neural network (CNN) and a corresponding classification strategy \cite{zhao2019multi}.
From the results of previous studies, deep learning can achieve more effective EEG feature extraction and higher precision classification. However, due to electrode displacements, skin-electrode impedance, different head shapes and sizes, different brain activity patterns, and disturbance by task-irrelevant brain activities, there are differences in EEG signals between different subjects or between different sessions \cite{cui2022eeg}. Because of the inter-subject variability, and also the non-stationarity of EEG signals. MI-based BCI usually needs a long calibration session for a new subject, from 20–30 min to hours or even days. This lengthy calibration significantly reduces the utility of BCI systems \cite{wu2022transfer}, but transfer learning can effectively reduce the effects of inter-session and inter-subject variability. Therefore, it is imperative to develop an effective transfer learning scheme for enhancing the usability of BCI systems. The study demonstrates that the better framework is the combination of the CNN and "fine-tune" transfer model \cite{zhang2021hybrid,zhang2021eeg_adaptive}.

In this paper, we provide a Dynamic Domain Adaptation Based Deep Learning Network (DADLNet) for addressing the inter-subject and inter-session variability in MI-BCI. We replace traditional EEG with a 3D array and use 3D convolution to learn temporal and spatial features. For the purpose of our model can better capture spatial-temporal information, we add an attention method that spatially combines convolutional channels. Furthermore, we develop a dynamic domain adaptation (DDA) strategy to adapt to the source domain in different scenarios, utilize maximum mean discrepancy (MMD) \cite{ZHANG2023126659} loss function to reduce the distance between the source and target domains to achieve the best results. We summarize the main contributions of our paper as:
\begin{itemize}
\item A MI classification network using an original feature extraction method combining temporal and spatial. The size of the temporal convolution kernel is related to the sampling rate of the input signal. Spatial feature extraction by local convolution maximizes preservation of event-related desynchronization (ERD)/event-related synchronization (ERS) features in left and right brain regions.
\item A spatial-channel attention is introduced to improve the DADLNet to decode the dependency between spatial and channel. 
\item A DDA strategy is proposed for both single-source and multi-source domains.
\end{itemize}

\section{Related Work}

\subsection{Attention Mechanism}

The attention mechanism is currently widely used in deep learning. Inspired by human visual observation, it has achieved remarkable achievements in various fields such as computer vision and natural language processing. Recently, many researchers have added attention mechanisms to analyze temporal and spatial correlations in MI recognition models. For instance, A spatial and temporal attention mechanism is proposed by \cite{LIU2022103001}. Temporal attention can improve the ability of the model to deal with the temporal non-stationarity of EEG signals. Spatial attention helps the network to focus on task-related EEG channels. Other authors \cite{9338374} have proposed a spatial and spectral attention mechanism to adaptively explore the most valuable information in EEG signals. An end-to-end spatiotemporal attention network (STAnet) \cite{su2022stanet} is proposed for detecting auditory spatial attention from EEG. STAnet dynamically assigns different weights to EEG signal channels through a spatial attention mechanism, and dynamically assigns different weights to temporal patterns of EEG signals through a temporal attention mechanism. An attention based convolutional recursive neural network (ACRNN) \cite{9204431} , which uses a channel attention mechanism to adaptively assign weights to different channels. In order to explore the temporal information of EEG signals, extended self attention is integrated into the RNN. However, the attention mentioned above does not take into account the correlation between channels generated by convolution. Based on this, we employ an attention method that combines spatial and inter-channel to capture the interaction between channels and spatial.

\subsection{Domain Adapation}

Domain Adapation (DA) has always been the most effective method in the field of deep learning to solve the variance between different objects. The proposal of Deep Adversarial Disentangled Autoencoder \cite{peng2019domain} provides an excellent solution for the target-agnostic DA field. In the face of multi-source domain scenarios (the data source is not unique), there will be a lot of invalid data that will have a negative impact on the model. Domain AggRegation Network \cite{wen2020domain} can eliminate the impact of invalid data in large samples and increase the effective sample size. It also can be applied in multiple realms of reality. For the diversity of source domains, there is the STEM \cite{nguyen2021stem} method that combines information from multiple source domains and perfectly handles the impact of source domain diversity on a single target domain. In recent years, many researchers have begun to use Deep Adaptation Network \cite{long2015learning} with MMD loss function for EEG domain adaptive research \cite{MMD_domain_adaptive}. A two-level domain adaptation neural network (TDANN) \cite{10.3389/fnhum.2020.605246} is proposed to construct a transfer model for EEG-based emotion recognition. First, a deep neural network is used for feature extraction, and then MMD and domain antagonism neural network (DANN) are used to bring the features closer to the corresponding class center. An auto-encoder architecture \cite{chai2016unsupervised} is proposed to reduce the discrepancy between training and test objects in the latent space. However, the above-mentioned DA method still lacks a simple and good solution when facing the problems of single-source domain and multi-source domain at the same time. Inspired by the MS-MDA framework \cite{chen2021ms}, we propose a DA strategy suitable for single-source domains and multi-source domains.

\section{Methods}

\subsection{3D Representation of EEG}

In this study, we use 3D temporal-spatial representations of EEG data as the input of the proposed DADLNet. There is variability in the EEG information acquired by electrodes in different regions on the electrode cap. Therefore, the data onto each electrode have spatial feature, but EEG-MI data was used to converted into a two-dimensional array (channel × time) in many research\cite{zhang2020motor,DynamicMI}, each column of the array represent the time point of EEG, and each row of the array is the channel of EEG. This method is helpful for networks learning the temporal feature, but loses the relative positional relationship between channels \cite{li2023mdtl}.
\begin{figure}[ht]
    \centering  
    \begin{subfigure}{0.9\linewidth}
        \centering
        \includegraphics[width=\linewidth]{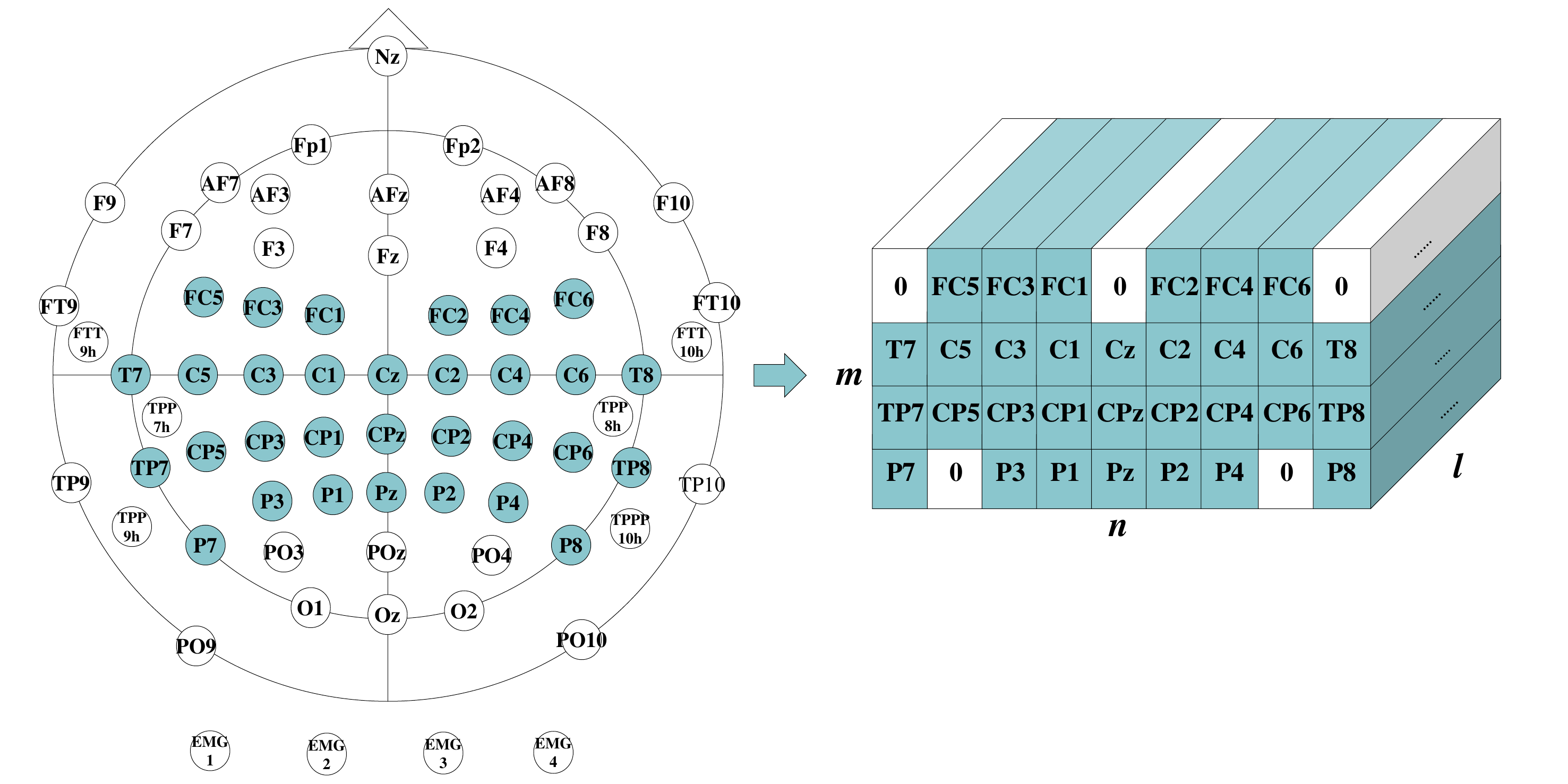}
        \caption{OpenBMI 3D representation}
        \label{3D_representation_a}
    \end{subfigure}

    \begin{subfigure}{0.9\linewidth}
        \centering
        \includegraphics[width=\linewidth]{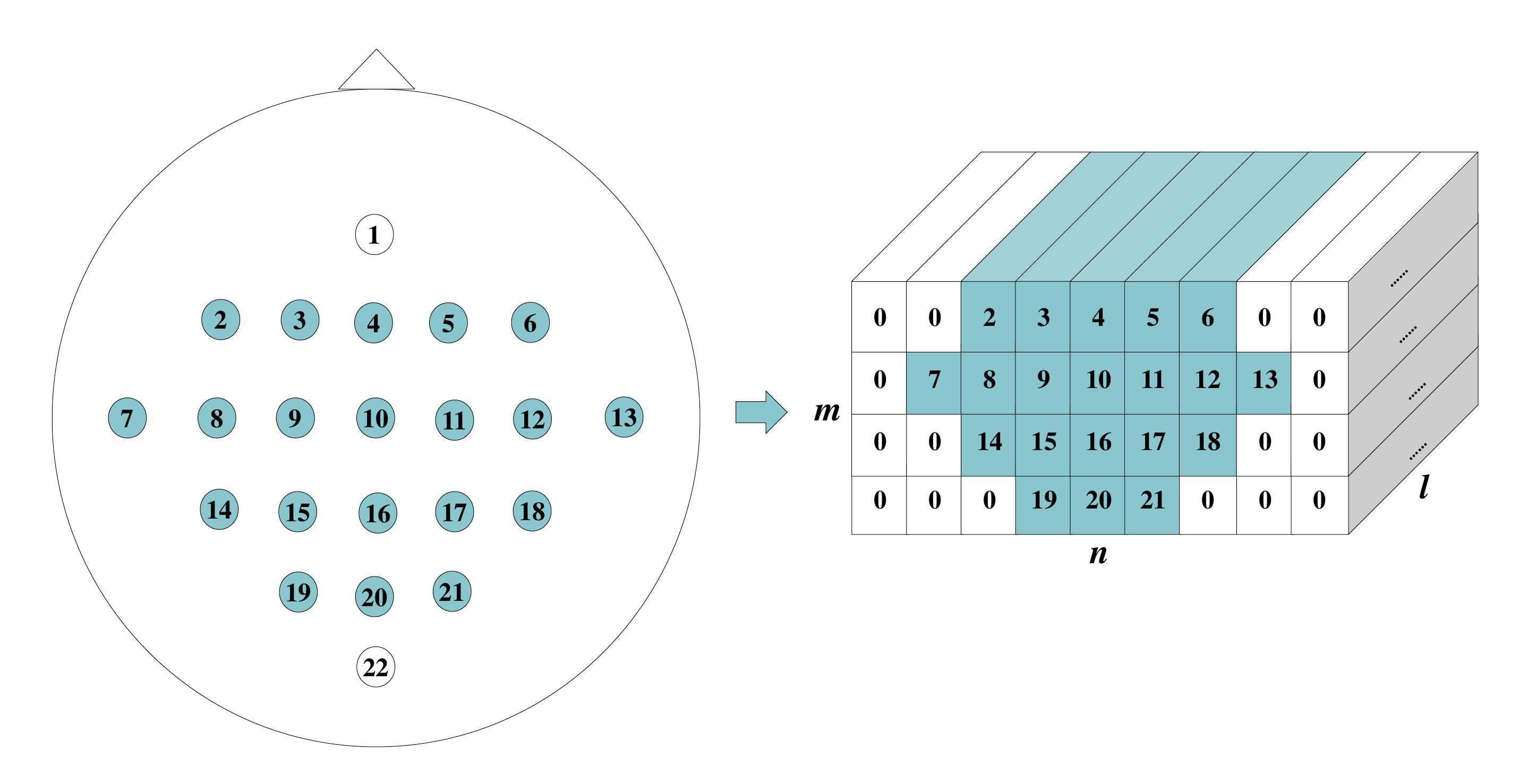}
        \caption{BCIC IV 2a 3D representation}
        \label{3D_representation_b}
    \end{subfigure}

    \caption{EEG 3D representation. UP: OpenBMI select 31 channels. Down: BCIC IV 2a select 20 channels. $l$ represents the Timesteps. $m$ and $n$ represents the EEG channel}
\end{figure}

To make full use of the temporal-spatial information of the EEG-MI data, we adopt the 3D representation of EEG data proposed in \cite{zhao2019multi}. First, We map the channels of the EEG data to the 3D array according to the electrode distribution shown in Fig. \ref{3D_representation_a} and Fig. \ref{3D_representation_b}, the first-dimension $l$ of the 3D array represents the sampling point of the EEG data, the second-dimension $n$ and the third-dimension $m$ correspond to the row and column of the channel respectively. The position in the array that does not have a corresponding channel is filled with 0. This 3D array representation not only preserves the temporal characteristics of the EEG data but also preserves the relative positional relationship of each channel.

\begin{figure*}[ht]
   \centering
    \includegraphics[width=\textwidth]{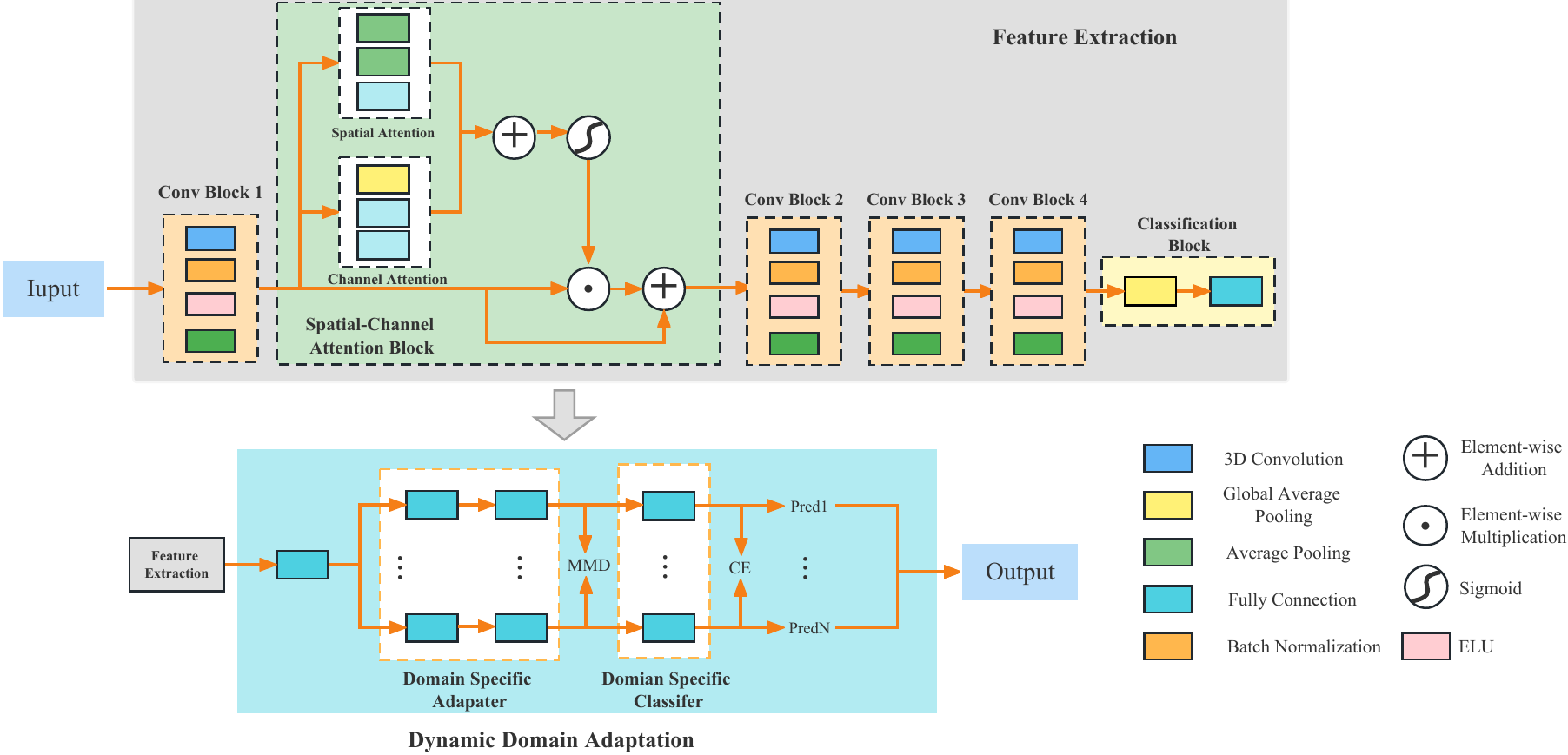}
    \caption{DADLNet architecture. All convolution operations are followed by batch normalization layer, ‘elu’ activation layer, average pooling layer, and dropout regularization in this order.}
    \label{figure1}
\end{figure*}

\subsection{Architecture of DADLNet}
The overall framework of our network is shown in Fig. \ref{figure1}, DADLNet consists of two main modules: Feature Extraction and Dynamic Domain Adaptation. The detailed structure of each module is explained in the following subsections.

\subsubsection{Feature Extraction}
The feature extraction module consists of three parts: Conv Block, Spatial-Channel Attention Block, and Classification Block. The Conv block consists of a 3D convolutional layer, a batch normalization (BN) layer, an exponential linear unit (ELU), and an average pooling layer.

\textbf{Conv Block:} A unique spatial convolution process is designed according to the spatial distribution of 3D representation data. The spatial dimension convolution kernel of the 3D convolutional layer in the Conv module is set to [(2,2),(2,2),(1,2),(2,2)], and the step size is set to [(1,1),(2,2),(1,2),(2,2)]. ERD/ERS phenomena occur in brain regions corresponding to the execution of left- or right-handed MI. So we use the smaller kernel size and step size to make the network learn the spatial features in the local range while ensuring the characteristics of the left and right brain regions were not mixed together. After the feature learning of the first three 3D convolution modules, the channel features are divided into four regions: "left middle part of the brain", "right middle part of the brain", "left lower part of the brain" and "right lower part of the brain". The last Conv module combines the learned local features. The feature maps extracted by the 3D convolutional layers in each Conv block are normalized by the BN layer. The ELU activation function plays the role of regularization and then downsamples the time dimension through the average pooling layer to reduce the amount of data while retaining effective information. Finally, the neurons are deactivated with a certain probability through the Dropout layer, which can not only reduce the calculation amount of the network but also improve the robustness of the model and prevent overfitting. In this paper, dropout = 0.5 is set.

\textbf{Spatial-Channel Attention Block:} The information carried by each channel in the EEG signal has different degrees of importance. Specifically, the same electrode signal also varies greatly for different people. Spatial attention can help the network to adaptively extract important information between channels. In addition, there is also a degree of importance between the feature maps generated by convolution in deep learning. Many channel attention mechanisms have been generated to greatly improve the accuracy of image classification in computer vision. Therefore, we use a combination of spatial and channel attention mechanisms in our network. Among them, the spatial attention \cite{jia2020sst} first uses channel-wise global average pooling to reduce computational costs. Next, using temporal-wise global average pooling to reduce the time dimension information. The final spatial attention matrix is computed by a fully connected (FC) layer. The channel attention first performs global average pooling for each channel and then generates attention weight parameters through two stages of Squeeze and Excitation (SE). The number of channels is squeezed using the reduction ratio parameter $r$ and the FC layer with Relu activation function into $C$$/$$r$ ($r$ = 8 in our model), and then the scaled vector back to its original shape. SE block has been proven to bring significant performance improvement at a small additional computing cost \cite{hu2018squeeze}. We combine the spatial attention module and SE block as part of our feature extraction. After calculating the spatial and channel attention values separately, we use element-wise addition to combine the feature maps and perform a sigmoid operation, followed by element-wise multiplication with the input data and appending to the output data. The specific modules are shown in Fig. \ref{figure1}.

\textbf{Classification Block:} The classification Block consists of a global average pooling layer and an FC layer. Different from traditional classifiers, we choose the global average pooling layer plus an FC layer as the final classification block, which greatly reduces the network parameters while retaining the information extracted by each convolutional layer and pooling layer. Finally, the probability value of each category is generated by a sigmoid.

\subsubsection{Dynamic Domain Adaptation}
Based on our DA strategy, since the experiment needs to face the problem of different number of source domains,we design the DDA, which consists of a FC layer, a Domain Specific Adapter (DSA) and a Domain Specific Classifier (DSC). First, we utilize a FC layer as a buffer layer, whose role is to map source and target domain data to a common feature space while extracting domain-invariant features in all domains. The structure of DDA is shown in Fig. \ref{figure1}.

\textbf{DSA:} It consists of two FC layers. In the 1-to-1 domain adapter, the data from the source domain and the target domain are mapped to a separate latent feature space. In this latent space, we employ MMD to measure the difference between the source domain and the target domain. The MMD function can then be defined as follows,
\begin{equation}
    \hspace{-0.5em} \operatorname{MMD}_{(X^S,X^T)} = \left\| \frac{1}{N^S} \sum_{i=1}^{N^S} \Phi(x_i^S) - \frac{1}{N^T} \sum_{j=1}^{N^T} \Phi(x_j^T) \right\|_{\mathcal{H}}^2,
    \label{eq:MMD}
\end{equation}
where $X^S$ and $X^T$ represent the source domain data and the target domain data respectively, $N^S$ and $N^T$ represent the vector lengths of $X^S$ and $X^T$ respectively, $\Phi$(.) representing a mapping function, and $H$ represents measurement by mapping both $X^S$ and $X^T$ to the regenerated Hilbert space, the value of MMD can be regarded as the distance between $X^S$ and $X^T$ in this space. By reducing the MMD loss, the difference between the source domain and the target domain is narrowed, so that the domain adapter can better predict the target domain features. 

\textbf{DSC:} It contains an FC layer, the activation function is sigmoid, and binary cross-entropy is used to measure the classification loss. DSC can make predictions about the target domain based on information from the current source domain. The specific definition of binary cross-entropy loss as
\begin{equation}
    \text{CE} = -\frac{1}{N} \sum_{i=1}^{N} y_{i} \cdot \log\left(p(y_{i})\right) + (1 - y_{i}) \cdot \log\left(1 - p(y_{i})\right),
\end{equation}
where $\mathrm{y}_{\mathrm{i}}$ represents the binary label 0 or 1, $p(.)$ represents the output probability of $\mathrm{y}_{\mathrm{i}}$, and $N$ represents the output dimension.

The characteristic of DDA is that it can be dynamically adjusted according to the number of source domains, so that the number of DSA and DSC is always consistent with the number of source domains. 

\subsection{Data Description}
To evaluate the effectiveness of the proposed model, we evaluate DADLNet and other baseline methods on OpenBMI \cite{lee2019OpenBMI} and BCIC IV 2a \cite{tangermann2012BCI2008}.

\textbf{OpenBMI:} This data set collects the EEG data of 54 healthy subjects (all right-handed, aged 24-35). EEG data were recorded using a device with 62 Ag/AgCl electrodes and a sampling rate of 1000 Hz. The experiment contains two sessions on different dates. Each session consists of an offline training phase and an online testing phase. Each phase has 100 binary classification MI tasks, and a total of 400 trials. In this paper, we chose 31 channels in the motor region (FC-5/3/1/2/4/6, T-7/8, C-5/3/1/2/4/6, Cz, TP-7/8, CP-5/3/1/z/2/4/6, P-7/3/1/z/2/4/8). The EEG signals were bandpass-filtered between 8Hz and 30Hz and downsampled from 1,000 to 400Hz. The MI segments from 0s to 4s after the stimulus started were selected for analysis.

\textbf{BCIC IV 2a:} This dataset collects the EEG data of 9 healthy subjects performing four categories (left hand, right hand, feet, tongue) MI tasks. EEG data were recorded using a device with 22 Ag/AgCl electrodes at a sampling frequency of 250 Hz. The dataset consists of two sessions on different dates, each session consists of 6 runs, each run contains 48 trials (performed 12 times for each type of MI task), and each session generates a total of 288 trials. In this paper, 20 channels (the selected channels are shown in Fig. \ref{3D_representation_b} are selected and only right-handed and left-handed MI task data were used. In order to ensure the same sampling frequency of the two datasets, the EEG data was upsampled from 250Hz to 400Hz, and the data were bandpass-filtered at 0.5Hz and 100Hz, and notch filtering at 50Hz to suppress line noise. Also select the 0s to 4s MI segment after stimulation for analysis.

\subsection{Baseline Models}
To verify the effectiveness of our proposed DADLNet, we compared it with MIN2Net \cite{autthasan2021min2net}, EEG-adapt \cite{zhang2021eeg_adaptive}, FBMSNet \cite{liu2022fbmsnet}, EEGNet \cite{lawhern2018eegnet}, and used the hyper-parameters described in the original paper.

\textbf{MIN2Net:} MIN2Nets combines deep metric learning with multi-task autoencoders to efficiently learn latent representations from EEG data. They compare the performance of the model intra and inter-subjects. the TensorFlow model and specific framework can be found at \cite{autthasan2021min2net}.

\textbf{EEG-adapt:} EEG-adapt is an adaptive network based on deep learning. By fine-tuning the pre-training model, it can learn the EEG features of the current subjects, effectively solving the problem of individual differences in EEG signals, and the experimental results verify the proposed the effectiveness of the method. The details can be found in \cite{zhang2021eeg_adaptive}.

\textbf{FBMSNet:} FBMSNet uses hybrid depth convolution to extract multi-scale temporal features, followed by a convolution module with a spatial filtering function and a variance layer that can extract temporal features. Finally, the introduced central loss function can effectively improve inter-subject compactness and inter-subject compactness. The related model and framework can be found in \cite{liu2022fbmsnet}.

\textbf{EEGNet:} EEGNet is a compact CNN that can be applied to multiple BCI paradigms and demonstrates its performance for both intra- and inter-subject classification. We used the same data preprocessing as in \cite{lawhern2018eegnet}, but the time window length of 2s.

\subsection{Experimental Evaluation}
Experimental evaluation consists of two phases, in the first phase we use the feature extraction module in Fig. \ref{figure1} to pre-train the model, and in the second phase we use the dynamic domain adaptive module to perform domain adaptive testing on the pre-trained model obtained in the first phase.

\subsubsection{Pre-training}
We evaluate the proposed method on the BCIC IV 2a and OpenBMI datasets both intra and inter-subject.

In intra-subject classification, the session1 data of the BCIC IV 2a dataset is divided into training set and validation set through 5-fold cross validation, and session2 is used as the test set. The non-feedback phase data of session1 and session2 of the OpenBMI dataset also use 5-fold cross-validation to re-divide the training set and validation set, and the feedback phase data of session1 and session2 are used to test the model.

In inter-subject classification, the data of the target subjects is used for the test set, and the data of the remaining subjects are randomly divided into a training set (80$\%$) and a validation set (20$\%$) through 5-fold cross validation.
 Because the sample number is small, this will lead to overfitting, so we adopted a sliding cut strategy to increase the number of samples. The size of the time window is consistent with the sampling rate, and the sliding step is 0.06 times the time window in the intra-subject experiments, and the step is set to 0.5 times the time window in the inter-subject experiments.

\subsubsection{Domain Adaptation Test}
In the classification in intra-subject, the session1 data and session2 data of BCIC IV 2a dataset are used as source domain data and target domain data respectively, and target domain data is divided into a fine-tuning training set and a test set through 5-fold cross-validation. The data are divided in steps of 0.06 times the time window, and there is no overlap between the fine-tuning training set and the test set for each fold. The feedback stage data of session1 and the feedback stage data of session2 of the OpenBMI dataset are used as target domain data. The rest is used as source domain data. The target domain data is also divided into a fine-tuning training set and a test set using 5-fold cross-validation. Considering the overall data volume of OpenBMI, we divide the data according to the step size of 1 (no sliding).

In inter-subject classification. The data of the target subject is used as the target domain data, and the data of the remaining subjects is used as the source domain data. The target domain data is divided into a fine-tuning training set and test set through 5-fold cross-validation. All data is divided according to the step size of 0.06.

In this part, we freeze the parameters of the feature extraction module. The experimental process is divided into three stages. The first stage uses labeled source domain data for classification task training. The second stage uses source domain data and target domain data for domain difference task training. The third stage involves five-fold cross-validation and testing phase. We accumulated and averaged the results of each fold to obtain the corresponding Accuracy, Specificity, Sensitivity, and f1-score.
\subsubsection{Train Strategy}
The model propose in this study is implemented based on the TensorFlow framework, and the experiment is completed using an NVIDIA GeForce RTX 2080Ti GPU with 11 GB of memory. In order to reduce the time of model training, we adopted an early stop mechanism. When the validation loss of 30 consecutive epochs does not decrease, the model stops training. The optimizer chooses $nadam$ ($lr$ = 0,001, $beta$ = 0.9,0.999). We also use a multi-process training method to avoid the influence of the results of the previous fold on the subsequent training.

\section{Result}
\subsection{MI Classification Result}

\begin{table*}[ht]
\small
\caption{Classification Performance (mean $\pm$ std) in $\%$ for the intra-subject on OpenBMI and BCIC IV 2a compared to baseline methods.}
\label{tabel1}
\vskip 0.05in
\begin{sc}
\resizebox{\linewidth}{!}{
\begin{tabular}{lccccc}
\toprule
DATA SET&METHODS&ACC&SEN&SPE&F1-SCORE\\
\midrule
\multirow{4}*{OpenBMI}
&MIN2Net&59.83$\pm$13.40&67.60$\pm$16.31&52.50$\pm$17.78&62.20$\pm$13.28\\
&EEG-adapt&64.91$\pm$15.92&68.70$\pm$23.89&61.11$\pm$26.51&64.53$\pm$19.91\\
&FBMSNet&69.69$\pm$14.32& \textbf{71.64$\pm$22.56}&67.93$\pm$26.62&68.97$\pm$17.41\\
&EEGNet&64.48$\pm$16.03&67.62$\pm$18.71&61.34$\pm$20.37&64.01$\pm$17.69\\
&DADLNet(ours) &\textbf{70.42$\pm$12.44} &69.85$\pm$14.98 &\textbf{71.02$\pm$12.67} &\textbf{69.59$\pm$13.60}\\
\midrule
\multirow{4}*{BCIC IV 2a}
&MIN2Net&65.23$\pm$16.14&-&-&64.72$\pm$18.39\\
&EEG-adapt&64.51$\pm$13.88&71.42$\pm$18.30&57.59$\pm$18.16&65.19$\pm$15.64\\
&FBMSNet&\textbf{83.83$\pm$12.14}&\textbf{83.36$\pm$16.52}&\textbf{84.29$\pm$8.67}&\textbf{83.16$\pm$13.39}\\
&EEGNet&68.53$\pm$15.57&67.89$\pm$20.34&69.17$\pm$14.89&65.76$\pm$19.24\\
&DADLNet(ours)&73.91$\pm$11.28&71.79$\pm$12.69&76.09$\pm$10.64&73.17$\pm$11.91\\
\bottomrule
\end{tabular}
}
\end{sc}
\vskip -0.1in
\end{table*}

\begin{table*}[ht]
\small
\caption{Classification Performance (mean $\pm$ std) in $\%$ for the inter-subject on BCIC IV 2a compared to baseline methods.}
\label{tabel2}
\vskip 0.05in
\begin{sc}
\resizebox{\linewidth}{!}{
\begin{tabular}{lccccc}
\toprule
DATA SET&METHODS&ACC&SEN&SPE&F1-SCORE\\
\midrule
\multirow{4}*{BCIC IV 2a}
&MIN2Net&60.03$\pm$9.24&-&-&49.09$\pm$23.28\\
&EEG-adapt&\textbf{78.73$\pm$7.47}&\textbf{80.96$\pm$8.32}&\textbf{76.51$\pm$10.98}&\textbf{78.96$\pm$7.36}\\
&FBMSNet&64.01$\pm$7.72&59.00$\pm$26.54&69.03$\pm$28.76&55.72$\pm$17.20\\
&EEGNet&64.61$\pm$8.67&61.01$\pm$22.63&68.22$\pm$17.27&59.66$\pm$18.06\\
&DADLNet(ours)&67.88$\pm$7.27&64.15$\pm$10.96&71.52$\pm$5.52&66.19$\pm$9.07\\
\bottomrule
\end{tabular}
}
\end{sc}
\vskip -0.1in
\end{table*}

The experimental results of DADLNet and the baseline model on both datasets are presented in Table \ref{tabel1}. According to the findings presented in \ref{tabel1}, it is evident that within the context of the intra-subjects experiment conducted on the BCIC IV 2a dataset, the performance of DADLNet is ranked second, surpassed only by FBMSNet. As shown in Tabel \ref{tabel2} the inter-subjects experiment conducted on BCIC IV 2a, it was observed that DADLNet exhibited superior performance compared to FBMSNet across all metrics. Notably, there was a notable increase of 3.87\% in accuracy and 10.47\% in the f1-score. The superior performance of FBMSNet in intra-subject experiments may be attributed to its suitability for decoding MI signals within the same subject. Upon further examination of the performance of DADLNet on the OpenBMI dataset in relation to other baseline methods utilized in intra-subject experiments, it becomes evident that DADLNet exhibits superior performance in terms of accuracy, specificity, and f1-score. In comparison to FBMSNet, DADLNet demonstrates enhancements in both average accuracy and f1-score, with improvements of 0.73\% and 0.62\% respectively. In comparison to other conventional deep learning techniques such as EEGNet, the DADLNet model exhibited an improvement in accuracy and f1-score by 5.98\% and 5.58\% respectively. It is worth mentioning that the standard deviation of DADLNet exhibits a significantly reduced magnitude compared to other fundamental networks, thereby substantiating the resilience of our proposed approach.

\subsection{Transfer Learning}
We compared model performance on three strategies: dynamic domain adaptive (DDA), fine-tuning (FT), and non-transfer learning (NTF). As shown in Table \ref{tabel3}, in the intra-subject experiment of the OpenBMI dataset, compared with the NTF, the accuracy of the binary classification task is increased by 1.37\% after the FT is adopted, and the accuracy of the binary classification task is increased by 3.16\% after the DDA is adopted. In the intra-object experiment of the BCIC IV 2a dataset, compared with the NTF, the accuracy rate of the binary classification task decreased by 2.01\% after the FT is adopted, and the accuracy rate of the binary classification task increased by 2.05\% after the DDA is adopted. As shown in Table \ref{tabel4}, in the inter-subject experiment of the BCIC IV 2a dataset, compared with NTF, after adding the FT, the accuracy of the binary classification task increased by 3.35\%, and after adding DDA, the accuracy of the binary classification task rate increased by 6.85\%. In these three experiments, DDA showed the best performance in terms of accuracy, sensitivity, specificity, and f1-score. At the same time, it can be seen from the std of each indicator that the robustness of the DDA is also the best.

\begin{table*}[ht]
\small
\caption{Classification Performance (mean $\pm$ std) in $\%$ for the intra-subject experimental results for DDA, FT and NTF on OpenBMI and BCIC IV 2a.}
\label{tabel3}
\vskip 0.05in
\begin{sc}
\resizebox{\linewidth}{!}{
\begin{tabular}{lccccc}
\toprule
DATA SET&METHODS&ACC&SEN&SPE&F1-SCORE\\
\midrule
\multirow{3}*{OPENBMI}
&DDA&\textbf{70.42$\pm$12.44}&\textbf{69.85$\pm$14.98}&\textbf{71.02$\pm$12.67}&\textbf{69.59$\pm$13.60}\\
&FT&68.63$\pm$13.15&69.23$\pm$14.67&68.09$\pm$14.08&68.37$\pm$13.82\\
&NFT&67.26$\pm$14.49&68.71$\pm$14.98&65.82$\pm$12.67&67.14$\pm$13.60\\
\midrule
\multirow{3}*{BCIC IV 2A}
&DDA&\textbf{73.91$\pm$11.28}&\textbf{71.79$\pm$12.69}&\textbf{76.09$\pm$10.64}&\textbf{73.17$\pm$11.91}\\
&FT&69.85$\pm$13.38&70.87$\pm$12.96&68.85$\pm$14.39&69.75$\pm$13.69\\
&NTF&71.86$\pm$12.55&70.22$\pm$20.64&73.51$\pm$10.21&69.71$\pm$17.19\\
\bottomrule
\end{tabular}
}
\end{sc}
\vskip 0.1in 
\end{table*}

\begin{table}[ht]
\small
\caption{Classification Performance (mean $\pm$ std) in $\%$ for the inter-subject experimental results for DDA, FT and NTF on BCIC IV 2a.}
\label{tabel4}
\centering
\vskip 0.05in
\begin{sc}
\resizebox{\linewidth}{!}{
\begin{tabular}{lccccc}
\toprule
DATA SET&METHODS&ACC&SEN&SPE&F1-SCORE\\
\midrule
\multirow{3}*{BCIC IV 2A}
&DDA  &\textbf{67.88$\pm$7.27}  &\textbf{64.15$\pm$10.96} &\textbf{71.52$\pm$5.52}  &\textbf{66.19$\pm$9.07}\\
&FT   &64.38$\pm$9.16           &64.05$\pm$9.72           &64.72$\pm$10.51          &64.02$\pm$9.45\\
&NTF  &61.03$\pm$7.64           &60.57$\pm$15             &61.50$\pm$13.58          &59.06$\pm$10.81\\
\bottomrule
\end{tabular}
}
\end{sc}
\vskip 0.1in 
\end{table}

\begin{figure*}[ht]
   \centering
   \includegraphics[scale=0.35]{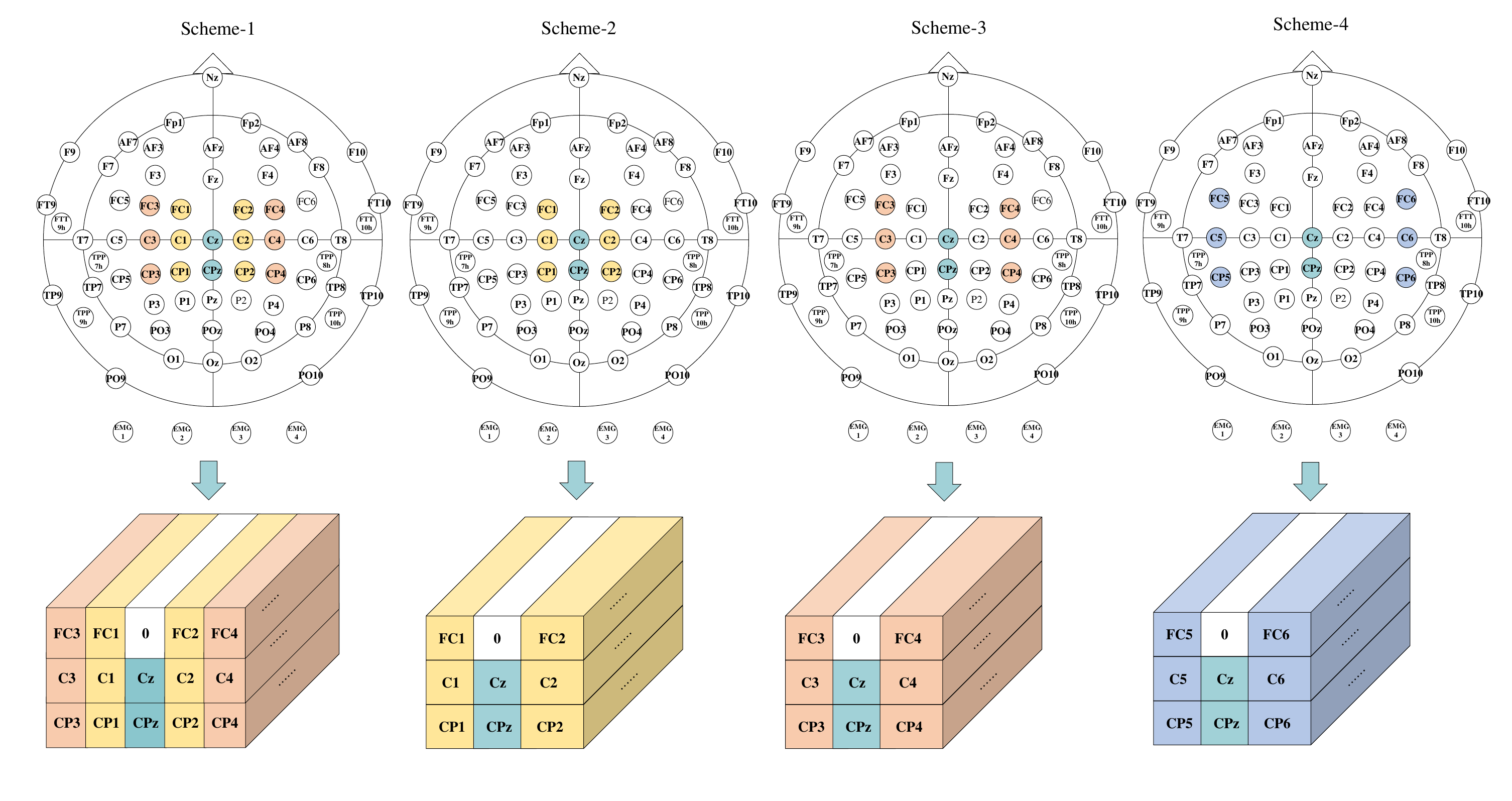}
   \caption{Four different channel selection schemes}
   \label{Discussion_channel}
\end{figure*}

\section{Discussion}
In this section, we discuss some of the experiments on the pre-trained model.
\subsection{Channel Select Strategy}
In real life, the number of electrodes is also a factor affecting the efficiency of BCI. Although the use of more electrodes can bring about an increase in accuracy, it will also lead to a longer preparation time for BCI. Therefore, this section explores our proposed framework in the selection classification performance with few channels.

As shown in Figure.\ref{Discussion_channel}, we have devised four different channel selection schemes. Scheme-1 chooses a total of 14 channels situated in three distinct regions: the anterior central zone (FC-1/2/3/4), the central zone (C-z/1/2/3/4), and the posterior central zone (CP-z/1/2/3/4). In the experimental setup, scheme-2, scheme-3, and scheme-4 were utilized to select a total of 8 channels for investigation. It is worth noting that Cz and CPz were identified as common channels across all three schemes. In addition to the common channels, Scheme-2 also has six additional channels, FC1, FC2, C1, C2, CP1, and CP2, which are primarily found near the midline of the brain and are better suited to capturing the majority of signals related to MI. Scheme-3 encompasses channels (FC3, FC4, C3, C4, CP3, CP4) that exhibit a slight displacement from the midline. However, it is noteworthy that these channels are situated within regions that bear greater significance to the cognitive process of MI specifically pertaining to the hand. Scheme-4 comprises channels (FC5, FC6, C5, C6, CP5, CP6) that are situated in proximity to the temporal lobe region. The aforementioned channels exhibited a higher degree of relevance in relation to the MI pertaining to the face. The primary objective of this study was to investigate the impact of electrode placement on the classification of the model. The network parameters and 3D data representation were adjusted in accordance with various electrode selection schemes.

\begin{table*}[ht]
\small
\caption{Comparison of results (mean $\pm$ std) of pre-trained models in binary classification intra-subject under different sampling electrode selection schemes.}
\label{tabel6}
\vskip 0.05in
\begin{sc}
\resizebox{\linewidth}{!}{
\begin{tabular}{lccccc}
\toprule
DATA SET&Scheme&ACC&SEN&SPE&F1-SCORE\\
\midrule
\multirow{5}*{OpenBMI}
&31-Channel &\textbf{67.26$\pm$14.49}&68.71$\pm$15.90&\textbf{65.82$\pm$18.03}&67.14$\pm$15.04\\
&scheme-1    &66.86$\pm$14.41&\textbf{69.08$\pm$14.97}&64.65$\pm$18.75&\textbf{67.26$\pm$14.35}\\
&scheme-2   &64.08$\pm$14.06&67.02$\pm$16.33&61.14$\pm$18.78&64.59$\pm$14.38\\
&scheme-3    &65.61$\pm$14.16&67.85$\pm$14.83&63.37$\pm$18.39&66.06$\pm$14.06\\
&scheme-4    &60.49$\pm$11.28&63.30$\pm$12.52&57.67$\pm$14.51&61.08$\pm$11.37\\
\bottomrule
\end{tabular}
}
\end{sc}
\vskip -0.1in
\end{table*}

Table \ref{tabel6} shows the results obtained by the different schemes on the OpenBMI dataset using 5-fold cross-validation combined with the leave-one-out validation method. Comparing the 31-channel and other channel selection settings, it can be observed that the average classification accuracy of the 31-channel is significantly better than that of scheme-2, scheme-3, and scheme-4, with a maximum improvement of 6.77\%. The average classification accuracy of scheme-1 is slightly lower than that of the 31-channel scheme, with a difference of 0.4\%, but the f1-score of scheme-1 is improved by 0.12\% compared to that of the 31-channel. This shows that our proposed method can achieve better results even when only a few channels are used. In addition, comparing scheme-2, 3, and 4, it can be found that the model performance of scheme-3 is better, followed by scheme-2, and the model performance of scheme-4 is the worst. This suggests that choosing different locations for the electrodes can have an impact on the model's performance with the same number of electrodes, and as mentioned earlier, the channel chosen for scheme-4 is mainly located in the lateral region of the brain, which can provide additional information about the face. However, the contribution of scheme-5 to the classification of right- and left-handed MI is not significant compared to the channels selected in scheme-2 and scheme-3. The channels selected in scheme-4 are particularly sensitive for capturing cortical activity related to hand movements and MI, and most of them are located close to the central sulcus of the brain. For the classification of right- and left-handed MI, these channels are considered to be the most critical and sensitive. Therefore, in terms of effectiveness, scheme-3 would be superior to scheme-2 and scheme-4. This further suggests that there are differences in the EEG signals recorded from electrodes in different regions, and hence the spatial characteristics of the MI EEG signals.

\subsection{The Influences of Time Kernel Size}
Studies have shown that setting the kernel size of the time dimension to the sampling rate [0.25,0.125] can extract frequency information above 4Hz and above 8Hz \cite{ding2020tsception}, and the ERD/ERS phenomenon related to MI mainly occurs in the mu (8-12Hz) and beta rhythm (18-26Hz), among which the change of mu rhythm is the most significant \cite{das2023hierarchical}. Therefore, in order to explore which time kernel size can better extract rhythm features related to MI, we designed the following four schemes (Shceme-A\footnote{0.25 times the sampling rate}, Shceme-B\footnote{0.125 times the sampling rate}, scheme-C\footnote{0.0625 times the sampling rate}, scheme-D\footnote{0.003125 times the sampling rate}). The kernel lengths of the time dimension are [0.25,0.125,0.0625,0.03125] of the sampling rate. According to \cite{bhattacharyya2017f1-score}, we choose F1-score as the evaluation metrics of the model.

\begin{table}[ht]
\small
\caption{Comparison of the F1-score of the pre-trained model in the binary classification intra-subject under different time-domain kernel sizes.}
\label{tabel7}
\vskip 0.05in
\begin{sc}
\resizebox{\linewidth}{!}{
\begin{tabular}{lcccc}
\toprule
Subject&Scheme-A&Scheme-B&Scheme-C&Scheme-D\\
\midrule  
1       &74.83 &72.98  &74.20  &\textbf{77.56} \\
2       &\textbf{52.69} &32.80  &30.08  &30.75\\
3       &86.85 &87.53  &\textbf{88.69}  &87.58\\
4       &57.50 &54.12  &55.00  &\textbf{58.56}\\
5       &60.50 &\textbf{68.02}  &54.05  &23.12\\
6       &60.01 &\textbf{60.78}  &57.45  &59.94\\
7       &68.41 &\textbf{78.29}  &74.95  &74.92\\
8       &84.80 &88.88  &\textbf{89.57}  &89.25\\
9       &73.56 &\textbf{83.95}  &74.21  &74.72\\
Mean    &68.79 &\textbf{69.71}  &66.47 &64.04\\
Stand   &11.42 &17.19   &17.98  &22.20\\
\bottomrule
\end{tabular}
}
\end{sc}
\vskip -0.1in
\end{table}
From the Table \ref{tabel7}, it can be observed that the average F1-score of scheme B is significantly better than that of scheme-A, C, and D. The model of scheme-A can also achieve a better F1-score value, and the model of scheme-D has the worst result. This shows that when the kernel size = 0.125, the model can effectively capture neurophysiological signals related to MI that appear in mu rhythm and beta rhythm. Therefore, this paper chooses scheme-B as the experimental setup. One of the reasons for obtaining suboptimal results with kernel size = 0.25 may be the interference of irrelevant information (4Hz-8Hz). When the kernel size is equal to 0.0625 and 0.03125 of the sampling rate, the model can only obtain part of the frequency information (16Hz and 32Hz) and the time domain receptive field of the model is small. These two reasons lead to the poor classification effect of the model.

Finally, comparing the performance of the same subject under different schemes, it can be found that the parameter settings for each subject to obtain the optimal performance are not the same, which further shows that there are individual differences in the MI EEG frequency segment. Therefore, a network design that can extract multiple frequency bands may achieve better performance.

\subsection{Future Work}
Although our proposed DADLNet achieves promising classification results, there is still room for further improvement. Firstly, the 3D representation of EEG can effectively preserve the spatial-temporal features of the data, but it lacks attention to the frequency domain characteristics. Some current studies have shown that considering the temporal-spatial-frequency domain information simultaneously can obtain good MI classification performance \cite{zhao2019temporal_spatial_frequency_1,9338374}. Secondly, DADLNet uses a fixed convolution kernel size to extract temporal information in the time dimension, but the optimal convolution kernel setting may vary from person to person \cite{dai2020fix_kernel_size}. Therefore, future work will consider combining multi-scale or multi-branch structures to further improve the classification performance of the model. Finally, we consider applying the model to the field of stroke rehabilitation in our future work to explore the possibility of DADLNet to extract MI features of stroke patients.

\section{Conclusion}
This paper presents DADLNet, a dynamic domain adaptation based deep learning network framework for decoding MI tasks. DADLNet consists of two parts: feature extraction module and dynamic domain adaptation module. The feature extraction part designs a feature extraction strategy that first learns local information and finally considers overall information, and innovatively combines the spatial and channel attention mechanisms to better extract discriminative information. The Dynamic domain adaptation module is uniquely designed to be suitable for both intra-subject as well as inter-subject experiments. Finally, the binary classification performance is compared with four advanced deep learning methods on the OpenBMI and BCIC IV 2a datasets. Experimental results verify the superiority of our proposed method.

\section*{References}
\bibliographystyle{IEEEtran}
\bibliography{manuscript}

\end{document}